\documentclass[aps,prb,reprint,showpacs,superscriptaddress]{revtex4-2}

\usepackage{amsmath}
\usepackage{amssymb}
\usepackage{mathrsfs}
\usepackage{graphicx}
\usepackage[colorlinks=true, linkcolor=blue, citecolor=blue, urlcolor=blue, pdfborder={0 0 0}]{hyperref}
\usepackage{setspace}
\usepackage{braket}
\usepackage{xcolor}  
\usepackage{bm}      

\newcommand{\ex}{\hat{e}_x}
\newcommand{\ey}{\hat{e}_y}
\newcommand{\Cv}{C_{\rm v}}
\newcommand{\mcO}{\mathcal{O}}

\graphicspath{{Figures/}}

\begin{document} 

\title{Subsystem symmetries, critical Bose surface, and immobile excitations in an extended compass model}

\author{Zhidan Li}
\thanks{These authors contributed equally to this work.}
\affiliation{College of Physics and Optoelectronic Engineering, Shenzhen University, Shenzhen 518061, China}
\affiliation{Institute of Physics, Chinese Academy of Science, Beijing 100872, China}

\author{Chun-Jiong Huang}
\thanks{These authors contributed equally to this work.}
\affiliation{Department of Physics and HKU-UCAS Joint Institute for Theoretical and Computational Physics at Hong Kong, The University of Hong Kong, Hong Kong, China}

\author{Hai-Zhou Lu}
\email{luhz@sustech.edu.cn}
\affiliation{Institute for Quantum Science and Engineering and Department of Physics, Southern University of Science and Technology, Shenzhen 518055, China}
\affiliation{Shenzhen Key Laboratory of Quantum Science and Engineering, Shenzhen 518055, China}

\author{Changle Liu}
\email{liuchangle89@gmail.com}
\affiliation{School of Engineering, Dali University, Dali, Yunnan 671003, China}
\affiliation{Institute for Quantum Science and Engineering and Department of Physics, Southern University of Science and Technology, Shenzhen 518055, China}
\affiliation{Department of Physics, and Center of Quantum Materials and Devices, Chongqing University, Chongqing 401331, China}
\date{\today}

\begin{abstract}
Subsystem symmetries are of peculiar interest as their relation to many exotic phenomena. However, realistic models hosting such symmetries are rare.
Here we propose an extended compass model that hosts subsystem symmetries and has potential experimental relevance to $3d$ transition metal compounds.
The subsystem symmetries strongly constrain the mobility of spin excitations and lead to profound consequences.
At the quantum critical point, we find the presence of a ``critical Bose surface'' along the entire $k_x$ and $k_y$ axes, and across this quantum critical point, we find a nodal-line spin liquid that undergoes nematic instability at low temperatures. 
In the ferroquadrupole phase, we find that one excitation is immobile individually analogous to ``fractons''. We discuss the relevance of our model to transition metal compounds.

\end{abstract}
\maketitle

\section{Introduction}
Symmetries lie at the heart of the fundamental principles in condensed matter physics.
For example, global symmetries play an essential role in classification of matters and critical behaviors within and beyond the Landau paradigm~\cite{Ref_SPT_00,Ref_SPT_01,Ref_DQCP_01,Ref_DQCP_02,Ref_SET_00,Ref_SET_01,Ref_SET_02,Ref_SET_03}, while local symmetries are responsible for various emergent gauge structures with fractionalization in spin liquids~\cite{Ref_SL_00,Ref_SL_01,Ref_SL_02,Ref_SL_03,Ref_SL_04,Ref_SL_05,Ref_SL_06} and fractional quantum Hall systems~\cite{Ref_FQHE_00,Ref_FQHE_01}. 
Recently, there has been intense interest in symmetries that interpolate between global and local ones. These symmetries are called ``subsystem symmetries'' (or ``quasi-local symmetries''), where symmetry operations are implemented only on subsets of the system~\cite{Ref_BoseMetal_00,Ref_CompassModel_22}. A well-known example is the ``Bose metal''~\cite{Ref_BoseMetal_00,Ref_BoseMetal_01,Ref_BoseMetal_02,Ref_BoseMetal_Mahan,Ref_BoseMetal_HanSangEun}, 
that preserves $U(1)$ boson number conservation within each row and each column.
These subsystem symmetries strongly constrain the boson dynamics, resulting a peculiar critical phase where bosons are neither gapped nor condensed. 
More generally, subsystem symmetries have been shown to be indispensable in fracton topological orders
~\cite{Ref_FracTopoOrder_00,Ref_FracTopoOrder_01,Ref_FracTopoOrder_02,Ref_FracTopoOrder_03,Ref_FracTopoOrder_04,Ref_FracTopoOrder_05,Ref_FracTopoOrder_06,Ref_FracTopoOrder_07,Ref_FracTopoOrder_08,Ref_FracTopoOrder_09,Ref_FracTopoOrder_10,Ref_FracTopoOrder_11,Ref_FracTopoOrder_12,Ref_FracTopoOrder_13,Ref_FracTopoOrder_14} and certain higher-order symmetry-protected topological phases
~\cite{Ref_SSPTO_00,Ref_SSPTO_01}.
More intriguingly, they lead to exotic physical behaviors such as dimensional reduction~\cite{Ref_DimensionalReduction_00} and UV-IR mixing~\cite{Ref_UVIRMixing_00,Ref_UVIRMixing_01} 
that even challenge the conventional renormalization group paradigm.
However, concrete microscopic models with such symmetries are not common, and most of them contain multiple spin interactions
~\cite{Ref_MultiSpinInteractions_00,Ref_MultiSpinInteractions_01,Ref_MultiSpinInteractions_02,Ref_MultiSpinInteractions_03}, which make them difficult to realize in experiments.

\begin{figure}[!t]
 \includegraphics[width=1.0\linewidth]{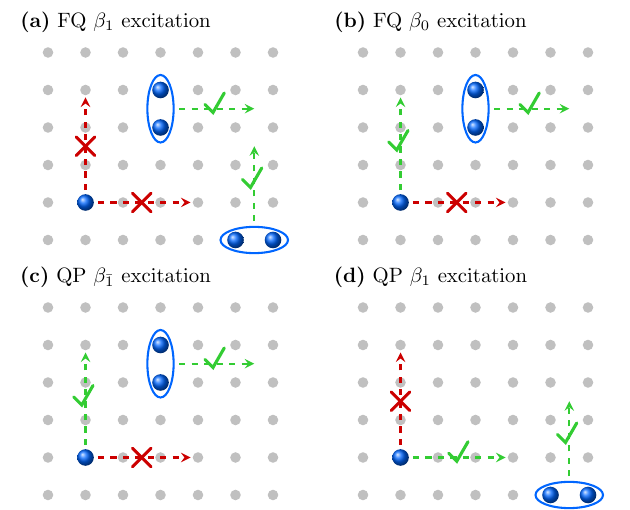}
 \caption{Mobility of excitations within (a)(b) the ferroquadrupole and (c)(d) the quantum paramagnetic phase. In particular, the immobility of the $\beta_1$ excitation in the ferroquadrupolar phase (a) resembles ``fractons''. Here the ellipse with two excitations only indicates that two flavor-wave excitations living in the same row (or column) that can cooperatively move along the transverse direction, and do not imply any physical bound state.}
 \label{fig:1}
\end{figure}

In this paper we propose an extended compass model
~\cite{Ref_CompassModel_00,Ref_CompassModel_01,Ref_CompassModel_02,Ref_CompassModel_03,Ref_CompassModel_04,Ref_CompassModel_05,Ref_CompassModel_06,Ref_CompassModel_07,Ref_CompassModel_08,Ref_CompassModel_09,Ref_CompassModel_10,Ref_CompassModel_11,Ref_CompassModel_12,Ref_CompassModel_13,Ref_CompassModel_14,Ref_CompassModel_15,Ref_CompassModel_16,Ref_CompassModel_17,Ref_CompassModel_18,Ref_CompassModel_19,Ref_CompassModel_20,Ref_CompassModel_21,Ref_CompassModel_22,Ref_CompassModel_23,Ref_CompassModel_24} 
that hosts subsystem symmetries within each row and column. This model only contains bilinear spin interactions and single-ion anisotropy, and is potentially relevant with $3d$ transition metal compounds. We demonstrate that these subsystem symmetries strongly constrain the quantum dynamics and impose profound physical consequences: At the quantum critical point, the system exhibits  ``critical Bose surface'' excitations located along the entire $k_x$ and $k_y$ axes in the reciprocal space. 
The nodal-line degeneracy is protected by one-dimensional subsystem symmetries hence cannot be broken at any finite temperatures~\cite{Ref_DimensionalReduction_00}. 
Across the transition, we find a peculiar liquid phase with the spin structural factor peaked along the entire $k_x$ and $k_y$ axes, which we dubbed as ``nodal-line spin liquid''. At low temperatures, the strong spin fluctuations further lead to nematic instabilities via order-by-disorder mechanism~\cite{Ref_OrderbyDisorder_00,Ref_OrderbyDisorder_01,Ref_OrderbyDisorder_02,Ref_OrderbyDisorder_03,Ref_OrderbyDisorder_04,Ref_OrderbyDisorder_05}. 
Besides, in the ferroquadrupole phase, we find that a branch of excitation, which is initially completely immobile individually, becomes mobile once such two excitations move simultaneously. 
This phenomenon is akin to that of ``fractons''. 
The symmetry-imposed immobility of spin excitations within the ferroquadrupole and quantum paramagnet phase is presented in Fig.~\ref{fig:1}.

The rest of the paper is arranged as follows.
In Sec.~\ref{sec_model}, we first introduce the extended compass model and its symmetries.
In Sec.~\ref{sec_phasediagram}, we establish the full phase diagram of the model through the semiclassical Monte Carlo method.
In Sec.~\ref{sec_mobility_Bose}, utilizing the linear flavor-wave theory, we elaborate on several peculiar physical phenomena observed in this model: restricted mobility excitations, critical Bose surface, nodal-line spin liquid, and fracton-like excitations.
Finally, we discuss the relevance of our model to transition metal oxide systems in Sec.~\ref{sec_discussions}.

\section{Extended compass model} \label{sec_model}
We propose an extended spin-1 compass model on a square lattice
\begin{equation}
    \mathcal{H} =  \sum_{\mathbf{r}} \big[  J (   S_{\mathbf{r}}^{x} S_{\mathbf{r}+\hat{e}_x}^{x}
    + S_{\mathbf{r}}^{y} S_{\mathbf{r}+\hat{e}_y}^{y}) 
    - D (S_{\mathbf{r}}^{z})^2 \big]
\label{Eq:ham}
\end{equation}
where $S_{\mathbf{r}}^{\alpha}$~($\alpha=x,y,z$) denotes the spin-1 operator at site $\mathbf{r}$, $J$ term represents the compass exchange coupling, and $D$ is the single-ion anisotropy. We find that the global $\mathbb Z_2$ operation $\mathcal G=\exp[{\sum_\mathbf{r} i(\mathbf{M} \cdot \mathbf{r}}) S_\mathbf{r}^z]$ with $\mathbf{M}=(\pi,\pi)$ changes the sign of $J$ while leaving $D$ invariant. Without loss of generality we set a ferromagnetic $J=-1$ and set $|J|$ as the energy unit. 

The extended compass model [Eq.~\eqref{Eq:ham}] hosts remarkable Ising subsystem symmetries defined on each row and column~\cite{Ref_CompassModel_10}: for each row $j$, we define $\mathcal P_j$ as $\pi$-rotation about the $y$ axis acting on this row,
\begin{equation}
    \mathcal P_j = \prod_{\mathbf{r}^{\prime} \in j} e^{-i\pi S_{\mathbf{r}^{\prime}}^{y}}.
\label{Eq:P_ham}
\end{equation}
Similarly, for each column $l$, we define $\mathcal Q_l$ as $\pi$-rotation about the $x$ axis acting on this column,
\begin{equation}
   \mathcal  Q_l = \prod_{\mathbf{r}^{\prime} \in l} e^{-i\pi S_{\mathbf{r}^{\prime}}^{x}}.
\label{Eq:Q_ham}
\end{equation}
Note that $\mathcal P$'s and $\mathcal Q$'s are Ising symmetries of the Hamiltonian  $[\mathcal P_j,\mathcal H]=[\mathcal Q_l,\mathcal H]=0$, $\mathcal P_j^2 =\mathcal Q_l^2= 1$, and are mutually commutative [$\mathcal P_j,\mathcal P_{j'}]=[\mathcal Q_l,\mathcal Q_{l'}]=[\mathcal P_j,\mathcal Q_l]=0$. 
Moreover, this system hosts time-reversal symmetry $\Theta$ and a spin-orbit-coupled $C_4$ rotational symmetry:
\begin{equation}
C_4: 
S_{\mathbf{r}}^{x}\rightarrow S_{\mathbf{r}^{~\!\!\prime}}^{y}, 
S_{\mathbf{r}}^{y}\rightarrow -S_{\mathbf{r}^{~\!\!\prime}}^{x},
S_{\mathbf{r}}^{z}\rightarrow S_{\mathbf{r}^{~\!\!\prime}}^{z},
\end{equation}
where  $\mathbf{r}'$ is the image of $\mathbf{r}$ under the $C_4$ rotation. 
Note that the subsystem symmetries [Eqs.~\eqref{Eq:P_ham} and \eqref{Eq:Q_ham}] were first discovered in Ref.~\cite{Ref_CompassModel_10} for the pure compass model $D=0$, and we point out that they still hold with finite $D$.
The presence of the $D$ term in Eq.~\eqref{Eq:ham} offers quantum tunability to the original compass model while keeping all the symmetries intact, hence one can keep track of the effects of subsystem symmetries by tuning this parameter.
In the next section, we construct the full phase diagram of this model, where $D$ varies from $-\infty$ to $+\infty$ and temperature ranges from zero to sufficiently high temperatures.

\section{Semiclassical phase diagram} \label{sec_phasediagram}
To establish the finite-temperature phase diagram of the extended compass model [Eq.~\eqref{Eq:ham}], we first employ the semiclassical Monte Carlo method (sMCM)~\cite{Ref_SemiClassical_00} accompanied by the parallel tempering. This semiclassical treatment is a powerful tool that can faithfully describe various quadrupole orders of spin-1 systems. 
In this section, we first describe the details of the sMCM, and then present the phase diagram obtained using this method.

\begin{figure}[!t]
\includegraphics[width=1.0\linewidth]{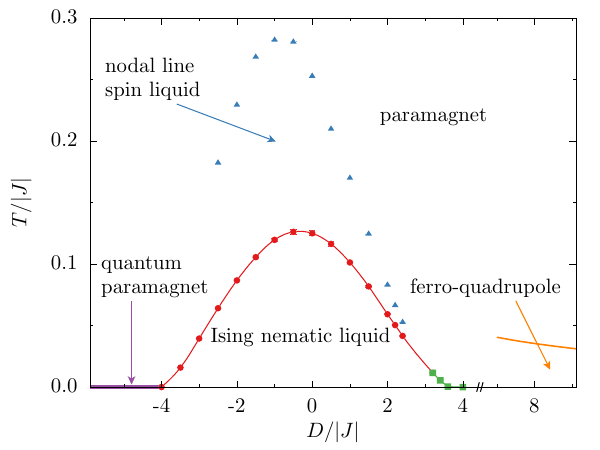}
\caption{
Semiclassical phase diagram of the extended compass model [Eq.~\eqref{Eq:ham}]. Red-solid lines correspond to continuous phase transitions while the green-solid lines correspond to the first-order transitions.
The blue-triangle points denote the crossover between nodal-line spin liquid and the paramagnet phase determined by the peak of the fluctuation $\chi_{\mathcal{O}_1}$ (see Appendices~\ref{app:B} and \ref{app:C}). 
The orange-solid line denotes the schematic thermal Ising phase transitions of the ferroquadrupole phase at large $D$ values, where this boundary is set to that of the effective classical Ising model [Eq.~\eqref{Eq:Heff2}], $T_c=\frac{1}{2\ln(1+\sqrt{2})}\frac{J^2}{2D}$. 
}
\label{fig:2}
\end{figure}

\subsection{The semiclassical Monte Carlo method} \label{sec_phasediagram_00}

Before we introduce sMCM, we first discuss the representation of a single $S=1$ spin. A general spin-1 state can be expressed as
\begin{equation}
    \ket{\mathbf{d}} = d_1\ket{1} + d_0\ket{0} + d_{\bar 1}\ket{\bar{1}},
    \label{eq:d}
\end{equation}
where $\ket{1},\ket{0},\ket{\bar{1}}$ are eigenstates of the operator $S^z$ with eigenvalues $1,0,-1$ respectively. Hence the state of a single spin can be represented by a three-dimensional complex vector $\mathbf{d}=(d_1,d_0,d_{\bar 1})$ satisfying the normalization constraint $\mathbf{d}^*\cdot\mathbf{d}=1$. 

In sMCM, we assume that the variational state of the many-body system is entanglement-free between different sites, hence take the direct product form,
\begin{equation}
    \ket{\Psi}=\otimes_\mathbf{r}|\mathbf{d}\rangle_\mathbf{r}
\end{equation}
where $\ket{\mathbf{d}}_\mathbf{r}$ is a state on site $\mathbf{r}$. Under this approximation, the expectation value of the Hamiltonian~\eqref{Eq:ham} can be computed as
\begin{equation}
    \begin{split}
    H &= \bra{\Psi}\mathcal{H}\ket{\Psi} \\
      &=  \sum_{\mathbf{r}} \left[ J (
        R_{\mathbf{r}}R_{\mathbf{r}+\ex} +
        I_{\mathbf{r}}I_{\mathbf{r}+\ey}
    ) - D (1-|d_{\mathbf{r},0}|^2) \right].
    \end{split}
\end{equation}
In this equation, $R_\mathbf{r}=\Re[d_{\mathbf{r},0}(d_{\mathbf{r},1}+d_{\mathbf{r},\bar 1})^*]$ and $I_\mathbf{r}=\Im[d_{\mathbf{r},0}(d_{\mathbf{r},1}-d_{\mathbf{r},\bar 1})^*]$ where $\Re[\cdot]$ and $\Im[\cdot]$ are the real and imaginary part of a complex number respectively. To address the finite-temperature properties, we rewrite the partition function as
\begin{equation}
    \begin{split}
    Z &= \textrm{Tr} \exp(-\beta\mathcal{H}) \sim \int \prod_{\mathbf{r}} \mathcal{D}_{\mathbf{d}_\mathbf{r}} \bra{\Psi} \exp(-\beta\mathcal{H}) \ket{\Psi} \\
    &\approx \int \prod_{\mathbf{r}} \mathcal{D}_{\mathbf{d}_\mathbf{r}} \exp\left[-\beta H\right],
    \end{split}
\end{equation}
where $\mathcal{D}_{\mathbf{d}_\mathbf{r}}=(2\pi)^2 \left[ \prod_{m=-1}^1 \mathrm{d} d_{\mathbf{r},m} \,\mathrm{d}d_{\mathbf{r},m}^*\right] \delta(\mathbf{d}_\mathbf{r}^*\cdot\mathbf{d}_\mathbf{r}-1)$ which means to perform an integration over the unit complex sphere in three dimensions. We do not encounter a sign problem because $H$ is real and a conventional Metropolis update can be implemented which updates $\ket{\mathbf{d}}_\mathbf{r}$ on site $\mathbf{r}$ according to the detailed balance equation. 
Note that there is an additional $U(1)$ gauge redundancy on each site: the $U(1)$ gauge transformations  $\mathbf{d}_{\mathbf{r}}\rightarrow \mathbf{d}_{\mathbf{r}} e^{i\theta_\mathbf{r}}$, where $\theta_\mathbf{r}$ are angular variables. This subtle redundancy, however, will not affect the sMCM simulations 
if we restrict our measurements to gauge-invariant observables. To overcome the ergodicity near the phase transition and at low temperatures, a general trick, the parallel tempering, is employed as well~\cite{Hukushima1996exchange}.
In Appendix~\ref{app:B}, based on the symmetries of the extended compass model, we give definitions of the observables measured in our simulations, such as the specific heat, Ising nematic order, and one-dimensional spin correlations, which are used to determine phases and phase boundaries.

\subsection{Semiclassical phase diagram} \label{sec_phasediagram_01}

\begin{figure*}[t]
  \includegraphics[width=1.00\textwidth]{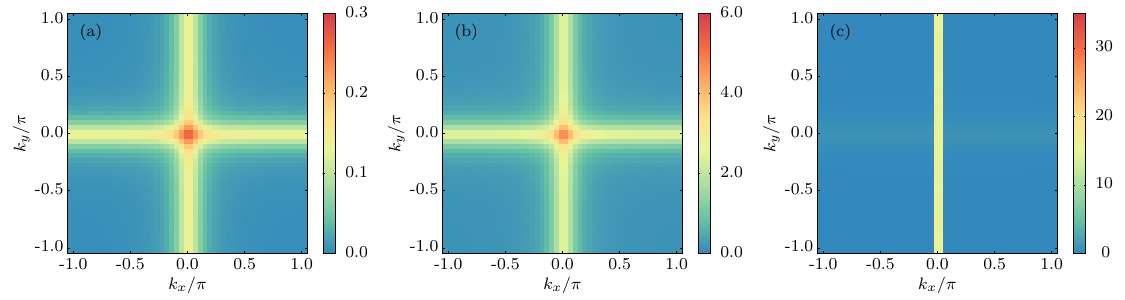}
 \caption{Spin structural factors measured
 (a) at the transition point $D=D_c$ with the temperature $T/|J|=0.01$, 
 (b) inside the nodal-line spin liquid, $(D, T/|J|)=(-0.5,0.2049)$,
 and
 (c) inside the Ising nematic phase, $(D, T/|J|)=(-0.5,0.1016)$. 
 The simulations are performed with linear size $L=48$ and (c) is measured in the symmetry breaking sector.
 }
 \label{fig:3}
\end{figure*}

Figure~\ref{fig:2} shows the finite-temperature phase diagram obtained through sMCM simulations. We begin by providing a qualitative discussion of the phase diagram at temperature $T\rightarrow 0$, followed by a more quantitative analysis of the finite-temperature phase diagram obtained from the sMCM simulations.
The qualitative phase diagram at $T\rightarrow 0$ reveals three distinct regimes: the regime where $D\rightarrow -\infty$, the regime where $D\rightarrow +\infty$, and the intermediate regime.
In the $D\rightarrow -\infty$ limit where single-ion $D$ term dominates, the system sits in the so-called ``large-$D$ quantum paramagnetic(QP)'' phase, a trivial product state of $S_\mathbf{r}^z=0$ at each site. 
The quantum paramagnetic state is protected by an energy gap $D$, hence it remains as the ground state with finite $J$.
In the $D\rightarrow +\infty$ limit, the semiclassical approximation fails to predict the correct ground state since it ignores quantum entanglement~(see Appendix~\ref{app:A}). However, degenerate perturbation theory predicts a twofold ferroquadrupole(FQ) order at low temperature. 
In the intermediate $D$ regime, we find a Ising nematic liquid
~\cite{Ref_CompassModel_09,Ref_CompassModel_15,Ref_CompassModel_16,Ref_CompassModel_21} which preserves time-reversal symmetry but breaks the spin-orbit-coupled $C_4$ symmetry down to $C_2$ [Fig.~\ref{fig:3}(c)]. The $C_4$ symmetry is restored upon increasing temperatures via a phase transition (the red and green lines in Fig.~\ref{fig:2}), and in a temperature window, we find the ``nodal-line spin liquid'' regime where the spin structural factors are sharply peaked along the entire $k_x$ and $k_y$ axes [Fig.~\ref{fig:3}(b)], much analogous to the spiral surface in ``spiral spin liquids.''~\cite{Ref_SpiralSpinLiquid_00,Ref_SpiralSpinLiquid_01,Ref_SpiralSpinLiquid_02,Ref_SpiralSpinLiquid_03,Ref_SpiralSpinLiquid_04,Ref_SpiralSpinLiquid_05,Ref_SpiralSpinLiquid_06,Ref_SpiralSpinLiquid_07,Ref_SpiralSpinLiquid_08,Ref_SpiralSpinLiquid_09,Ref_SpiralSpinLiquid_10}

\section{Restricted mobility of excitations and critical Bose surface}\label{sec_mobility_Bose}

To understand the implications of the subsystem symmetries, we investigate the spin excitations of the quantum paramagnetic phase and the ferroquadrupole phase with the flavor-wave theory
~\cite{Ref_FlavorWave_00,Ref_FlavorWave_01,Ref_FlavorWave_02}. 
In this section, we first present the details of the linear flavor-wave theory, and then reveal that there exist restricted mobility of excitations in the quantum paramagnetic phase, a nodal-line spin liquid in the intermediate regime and fracton-like excitations above the ferroquadrupole state. 

\subsection{Linear flavor-wave theory}

Here we present the details of the flavor-wave theory
~\cite{Ref_FlavorWave_00,Ref_FlavorWave_01,Ref_FlavorWave_02}. For spin-1 systems, the local basis of a spin at site $\mathbf{r}$ can be expressed in terms of three flavors of bosons,
\begin{equation}
    b_{\mathbf{r},m}^{\dagger} |\text{vac} \rangle \equiv | m \rangle_\mathbf{r},
\end{equation}
where $|\text{vac} \rangle$ indicates the bosonic vacuum, $| m \rangle_\mathbf{r}$ ( $m= 1, 0, -1$) are the eigenstates of $S_\mathbf{r}^z$ with eigenvalue $m$. The Hilbert space of the bosons is larger than the original spin Hilbert space and includes unphysical states. To limit the boson Hilbert space to its physical sector, a hard constraint must be imposed on each site $\mathbf{r}$,
\begin{equation}
    \sum_{m=-1}^{1} b_{\mathbf{r},m}^{\dagger} b_{\mathbf{r},m} = 1.
\end{equation}
Within the flavor-wave formalism, the relevant on-site spin operators can be expressed as quadratics of flavor bosons
\begin{equation}
\begin{split}
    S^z &= b_1^{\dagger}b_1 - b_{\bar{1}}^{\dagger}b_{\bar{1}},       \\
    (S^z)^2 &= b_1^{\dagger}b_1 + b_{\bar{1}}^{\dagger}b_{\bar{1}},       \\
    S^{+} &= \sqrt{2}~ (b_1^{\dagger}b_0 + b_{0}^{\dagger}b_{\bar{1}} ),  \\
    S^{-} &= \sqrt{2}~ (b_{\bar{1}}^{\dagger}b_0 + b_{0}^{\dagger}b_{1} ),
\end{split}
\end{equation}
where $S^{\pm}\equiv S^x \pm i S^y$. Here the site index $\mathbf{r}$ is omitted for simplicity. To simplify our discussions, we further define a rotated basis
\begin{equation}
    \begin{pmatrix}
    \beta_1 \\ \beta_0 \\ \beta_{\bar 1}
    \end{pmatrix}
    =\begin{pmatrix}
    \frac{1}{\sqrt{2}} & & \frac{1}{\sqrt{2}} \\
     & 1 & \\
     -\frac{1}{\sqrt{2}} & & \frac{1}{\sqrt{2}}
    \end{pmatrix}
    \begin{pmatrix}
    b_1 \\ b_0 \\ b_{\bar 1}
    \end{pmatrix}
\end{equation}
for the extended compass model.

In flavor-wave theory, different magnetic orders can be obtained by condensing the corresponding flavor bosons. In the following, we discuss the quantum paramagnetic and the ferroquadrupole phases, respectively.

\subsection{Restricted mobility of excitations in the quantum paramagnetic phase}

\begin{figure*}[!t]
  \includegraphics[width=1.0\textwidth]{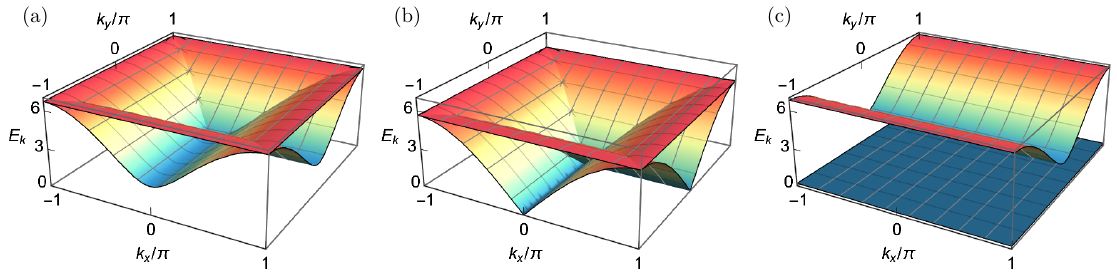}
 \caption{Dispersions of flavor-wave excitations (a)~in the quantum paramagnetic phase $D=-5.0$, (b)~at the quantum critical point $D_c=-4.0$, and (c)~in the ferroquadrupole phase $D=5.0$.
 }
 \label{fig:4}
\end{figure*}

In the quantum paramagnetic phase, the $\beta_0$ ($b_0$) flavor is condensed
\begin{align}
    \beta_0^{\dagger} = \beta_0 \approx \sqrt{1 - \beta_1^{\dagger}\beta_1 - \beta_{\bar{1}}^{\dagger}\beta_{\bar{1}} },
\end{align}
while the other two boson flavors $\beta_1$ and $\beta_{\bar 1}$ are viewed as the excitations above the quantum paramagnetic state. Rewriting the spin Hamiltonian of Eq.~\eqref{Eq:ham} in terms of the flavor bosons, and then expanding it in terms of $\beta_1$ and $\beta_{\bar 1}$ up to the quadratic order, and then performing the Fourier transformation, we obtain the linear flavor-wave Hamiltonian
\begin{align}
\mathcal H_{\mathrm{QP}} & \!=\! \frac{1}{2}\!\sum_{\mathbf{k}}\psi_{\mathbf{k},1}^\dagger\!\!
\begin{pmatrix}
-D+2J\cos k_{x} & 2J\cos k_{x}\\
2J\cos k_{x} & -D+2J\cos k_{x}
\end{pmatrix}\!\!
\psi_{\mathbf{k},1}
\nonumber\\
 & \!+\! \frac{1}{2}\!\sum_{\mathbf{k}}\psi_{\mathbf{k},\bar 1}^\dagger\!\!
\begin{pmatrix}
 -D+2J\cos k_{y} & -2J\cos k_{y}\\
-2J\cos k_{y} & -D+2J\cos k_{y}
\end{pmatrix}\!\!
\psi_{\mathbf{k},\bar 1},
\label{Eq:Ham_QP}
\end{align}
where we denote $\psi_{\mathbf{k},m}^\dagger=(\beta_{\mathbf{k},m}^{\dagger},\beta_{-\mathbf{k},m})$ for $m=\bar 1,1$. We find that in Eq.~\eqref{Eq:Ham_QP} the $\beta_1$ and $\beta_{\bar 1}$ branches are decoupled at the quadratic level for the reason that will be discussed later. The dispersions of the $\beta_1$ and $\beta_{\bar 1}$ excitations can be directly obtained from Bogoliubov transformation, $E_{\mathbf k,1}=\sqrt{D^2-4DJ\cos k_x}$ and $E_{\mathbf k,\bar 1}=\sqrt{D^2 -4DJ\cos k_y}$, see Fig.~\ref{fig:4}(a). 

The excitations acquire a gap $\Delta=\sqrt{D^2-4|DJ|}$, dictating the discrete symmetries of the model. For the ferromagnetic $J<0$ case, the band minima of the two modes are located at the entire $k_x=0$ and $k_y=0$ lines in the Brillouin zone, respectively, in contrast with usual models where the minimum locates only at some discrete points. 
Moreover, we note that both flavor-wave excitations become dispersionless along a particular direction, which indicates that the excitations are mobile only along one direction, and becomes immobile along another direction [Figs.~\ref{fig:1}(c)~and~\ref{fig:1}(d)]. 
We point out that this feature is not an artifact of the linear flavor-wave approximation, but deeply rooted in the subsystem symmetries of this system.

Here we derive the symmetry representations of the flavor-wave excitations under subsystem symmetries [Eqs.~\eqref{Eq:P_ham}~and~\eqref{Eq:Q_ham}]. All flavor-wave excitations have definite parities under subsystem symmetries [Eqs.~\eqref{Eq:P_ham}~and~\eqref{Eq:Q_ham}]. 
Due to the presence of the condensate $\beta_0$ in the quantum paramagnetic phase, the flavor-wave excitations can be approximated in the operator form $\beta_{1}\approx\beta_0^\dagger \beta_{1}=\frac{1}{2}\left(S^{x}-iQ^{yz}\right)$ and $\beta_{\bar{1}}\approx\beta_0^\dagger \beta_{\bar 1}=\frac{1}{2}\left(-Q^{xz}+iS^{y}\right)$. Here $Q^{xz}\equiv\{S^{x},S^{z}\}$ and $Q^{yz}\equiv\{S^{y},S^{z}\}$ correspond to on-site $xz$ and $yz$ spin quadrupole operators, respectively. It is straightforward to verify the following algebraic relation:
\begin{subequations}
\begin{align}
[\beta_{\mathbf{r},1},\mathcal P_j]&=[\beta_{\mathbf{r},\bar 1},\mathcal P_j]=0,   \, j\neq r_y, \label{Eq:QPa1}\\
[\beta_{\mathbf{r},1},\mathcal Q_l]&=[\beta_{\mathbf{r},\bar 1},\mathcal Q_l]=0, \, l\neq r_x, \label{Eq:QPa2}\\
\{\beta_{\mathbf{r},1},\mathcal P_{r_y}\}&=[\beta_{\mathbf{r},1},\mathcal Q_{r_x}]=0, \label{Eq:QPa3}\\
\{\beta_{\mathbf{r},\bar 1},\mathcal Q_{r_x}\}&=[\beta_{\mathbf{r},\bar 1},\mathcal P_{r_y}]=0. \label{Eq:QPa4}
\end{align}
\label{Eq:sym_QP}
\end{subequations}
From the above relation, we can see that the $\beta_{\mathbf{r},1}$ ($\beta_{\mathbf{r},\bar 1}$) excitation is even under all subsystem symmetries except the $\mathcal P$ at the same row $\mathcal P_{r_y}$ (the $\mathcal Q$ at the same column $\mathcal Q_{r_x}$).

Now the immobility nature of flavor-wave excitations becomes clear: 
Subsystem symmetries strongly constrain the linear mixing of flavor bosons, as only bosons carrying exactly the same symmetry representations are allowed to hop or pair.
We can see that all $\beta_1$ along the same row (and all $\beta_{\bar{1}}$ along the same column) carry exactly the same representation hence can be mixed linearly, while all other combinations are disallowed.
This is precisely reflected in the flavor-wave Hamiltonian [Eq.~\eqref{Eq:Ham_QP}]:
the $\beta_{1}$ excitation is mobile along the $x$ direction and becomes immobile along the $y$ direction; Similarly, the $\beta_{\bar 1}$ excitation is mobile along the $y$ direction and becomes immobile along the $x$ direction.
As a result, a single flavor-wave excitation is effective one-dimensional that can only propagate along one direction. However, a pair of $\beta_1$ excitations at the same row (or a pair of $\beta_{\bar 1}$ at the same column) commute with all subsystem symmetries hence can cooperatively propagate throughout the 2D plane, see Figs.~\ref{fig:1}(c)~and~\ref{fig:1}(d).

Although the previous derivation was based on the linear flavor-wave framework, our conclusion on the mobility of the flavor-wave excitations is deeply rooted in symmetries~\cite{Ref_CompassModel_22} that goes beyond the na\"ive linear flavor-wave approximation. 
In Appendix~\ref{app:D}, we will show that this conclusion still holds in the presence of flavor-wave interactions.

\subsection{Critical Bose surface and nodal-line spin liquid}
The symmetry protected immobility of excitations have profound implications on the nature of criticality and the proximate phase. 
Here we analyze the magnetic instabilities of the quantum paramagnetic state. 
As we turn on larger $|J|/D$ in the quantum paramagnetic phase, the bands become more dispersive. Until we reach a critical value of $D_c$ the excitations become gapless, and the transition occurs. 
At the quantum critical point $D_c$, the gapless modes constitute the ``critical Bose surface'' along the entire $k_x$ and $k_y$ axes~[Fig.~\ref{fig:4}(b)], and is protected by the subsystem symmetries as well as the $C_4$ symmetry. The existence of the Bose surface can be further illustrated by measuring the spin structural factor
\begin{equation}
    \mathcal S(\mathbf{Q})=\frac {1}{N}\sum_{\mathbf{rr}'}\langle \mathbf{S}_\mathbf{r}\cdot \mathbf{S}_{\mathbf{r}'} \rangle e^{i\mathbf{Q}\cdot(\mathbf{r}-\mathbf{r}')}.
\end{equation}
From Fig.~\ref{fig:3}(a) we see that $\mathcal S(\mathbf{Q})$ is clearly peaked along the entire $k_x$ and $k_y$ axes, consistent with the Bose surface scenario. The direction-dependent immobility renders the excitations 1D-like, and implies the specific heat scaling $C_v \sim T$ at low temperatures. More discussions about the nature of this transition will be given in the follow-up paper~\cite{unpublished}.

The intermediate phase can be understood from proliferation of $\beta_1$ and $\beta_{\bar 1}$ excitations in the quantum paramagnetic phase.
Due to the nodal-line degeneracy, the structural factor should be peaked along the entire $k_x$ and $k_y$ axes, which signifies absence of magnetic long-range order. In fact, the above scenario only holds at a finite temperature window as shown by the ``nodal-line spin liquid'' regime in Fig.~\ref{fig:2}, with spin structural factor shown in Fig.~\ref{fig:3}(b). Upon decreasing temperatures, the strong spin fluctuations spontaneously lift the degeneracy between the $\beta_1$ and $\beta_{\bar 1}$ bands and develop an Ising nematic order. The nematic order parameter takes the form
\begin{equation}
\hat{ \mathcal{O}}_N=\frac{1}{N}\sum_{\mathbf{r}} (S_{\mathbf{r}}^{x} S_{\mathbf{r}+\hat{e}_x}^{x}-S_{\mathbf{r}}^{y} S_{\mathbf{r}+\hat{e}_y}^{y})
\label{Eq:Nematic}
\end{equation}
that breaks the $C_4$ symmetry down to $C_2$. From the structural factor inside the nematic phase [Fig.~\ref{fig:3}(c)], we observe that spins are completely uncorrelated along the $x$ or $y$ direction, hence can be regarded as decoupled 1D chains.

\subsection{Fracton-like excitations above the ferroquadrupole state}
Here we discuss the ferroquadrupole phase, which can be well understood from the limit of $D\rightarrow +\infty$. In the large-$D$ limit the $S_\mathbf{r}^z=0$ state has a large energy penalty of $D$ and the low-energy subspace is spanned by $S_\mathbf{r}^z=\pm1$ states. One can thus define effective spin-1/2 operator $\tau_\mathbf{r}$ acting on the $S_\mathbf{r}^z=\pm1$ subspace
\begin{align}
    \tau_\mathbf{r}^z &= \frac{1}{2} \mathcal P_\mathbf{r} S_\mathbf{r}^z \mathcal P_\mathbf{r}, \\
    \tau_\mathbf{r}^{\pm} &= \frac{1}{2} \mathcal P_\mathbf{r} (S_\mathbf{r}^{\pm})^2 \mathcal P_\mathbf{r},
\end{align}
where $\mathcal P_\mathbf{r}$ is a projection operator onto the low-energy $S_\mathbf{r}^z=\pm1$ subspace.

The low-energy effective Hamiltonian can be obtained from second-order perturbation theory in the limit $D\gg |J|$. With straightforward calculations, we find that it turns out to be a ferromagnetic Ising model of the $\tau$ variables~(see Appendix~\ref{app:A}),
\begin{equation}
    \mathcal H_{\text{eff}}^{(2)} = -\frac{J^2}{2D} \sum_{\mathbf{r}}
    (\tau_{\mathbf{r}}^x \tau_{\mathbf{r}+\hat{e}_x}^x + \tau_{\mathbf{r}}^x \tau_{\mathbf{r}+\hat{e}_y}^x).
    \label{Eq:Heff2}
\end{equation}
Therefore, the ground state should be $\tau^x \sim (S^{x})^2-(S^{y})^2$ ferroquadrupole ordered. The twofold ferroquadrupole order breaks the same symmetry as the Ising nematic phase, but described by an on-site order parameter $\tau^x$. 

Here we discuss the flavor-wave excitations above the ferroquadrupole order. From the perturbation analysis in Appendix~\ref{app:A}, we can know that the ferroquadrupolar order arises purely from the quantum perturbative effect that involves mutual entanglement between different sites, which is ignored in the linear flavor-wave formalism. To obtain the correct result, such perturbative corrections [Eq.~\eqref{Eq:Heff2}] must be taken into account in the flavor-wave calculations.
Without losing generality we investigate the $\tau^x=-1/2$ ground state where
\begin{align}
    \beta_{\bar 1}^\dagger=\beta_{\bar 1}\approx\sqrt{1-\beta_{0}^{\dagger} \beta_0-\beta_1^\dagger \beta_1}
\end{align}
is condensed. With the second-order perturbative corrections [Eq.~\eqref{Eq:Heff2}] taken into account, the resulting linear flavor-wave Hamiltonian of ferroquadrupole order takes the form:
\begin{align}
\mathcal H_{\mathrm{FQ}} & =\frac{1}{2}\sum_{\mathbf{k}}
\psi_{\mathbf{k},0}^\dagger
\begin{pmatrix}
M(\mathbf{k}) & -2J\cos k_{y}\\
-2J\cos k_{y} & M(\mathbf{k})
\end{pmatrix}
\psi_{\mathbf{k},0}
\nonumber\\
 & +\frac{J^{2}}{D}\sum_{\mathbf{k}}
\beta_{\mathbf{k},1}^\dagger \beta_{\mathbf{k},1},
\label{Eq:Ham_FQ}
\end{align}
where $\psi_{\mathbf{k},0}^\dagger=(\beta_{\mathbf{k},0}^{\dagger},\beta_{-\mathbf{k},0})$ and $M(\mathbf{k})=D+\frac{J^{2}}{2D}+2J\cos k_{y}$.
From the energy dispersions [Fig.~\ref{fig:4}(c)], we find that the excitations are spatially anisotropic, reflecting the nematic nature of the ferroquadrupole phase. 

Surprisingly, we find that the $\beta_1$ band is completely flat, indicating that a single $\beta_1$ excitation is completely immobile individually.
To understand the symmetry imposed mobility constraint, we analyze the parities of the flavor-wave excitations under subsystem symmetries.  For the $\tau^x =-1/2$ state where $\beta_{\bar 1}$ is condensed, $\beta_1 \approx \beta_{\bar 1}^\dagger \beta_1=-\frac{1}{2} (S^z+iQ^{xy})$ and $\beta_0 \approx \beta_{\bar 1}^\dagger \beta_0=-\frac{1}{2}(Q^{xz}+iS^y)$. Here $Q^{xy}\equiv \{S^x,S^y\}$ is the $xy$-type spin quadrupole operator. 
The algebraic relation becomes the following:
\begin{subequations}
\begin{align}
[\beta_{\mathbf{r},1},\mathcal P_j]&=[\beta_{\mathbf{r},0},\mathcal P_j]=0,   \, j\neq r_y, \label{Eq:FQpart1} \\
[\beta_{\mathbf{r},1},\mathcal Q_l]&=[\beta_{\mathbf{r},0},\mathcal Q_l]=0, \, l\neq r_x, \label{Eq:FQpart2} \\
\{\beta_{\mathbf{r},1},\mathcal P_{r_y}\}&=\{\beta_{\mathbf{r},1},\mathcal Q_{r_x}\}=0, \label{Eq:FQpart3} \\
[\beta_{\mathbf{r},0},\mathcal P_{r_y}]&=\{\beta_{\mathbf{r},0},\mathcal Q_{r_x}\}=0. 
\label{Eq:FQpart4}
\end{align}
\label{Eq:sym_FQ}
\end{subequations}
\noindent We find that a single $\beta_{\mathbf{r},1}$ excitation is odd under the $\mathcal P$ the same row $\mathcal P_{r_y}$ and the $\mathcal Q$ at the same column $\mathcal Q_{r_x}$, while it commutes with all other subsystem symmetries. This means that the $\beta_1$ excitations at different sites carry different representation under subsystem symmetries [Eqs.~\eqref{Eq:P_ham}~and~\eqref{Eq:Q_ham}], hence could not hop or pair between different sites and become completely immobile.
However, a pair of $\beta_1$ excitations at the same row (or at the same column) can propagate along the direction transverse to the row (or column). The mobility of the $\beta_1$ excitations are much analogous to the ``type-I fractons''. 
Notwithstanding, it is different from fractons since it belongs to non-topological excitations that can be created/annihilated individually.

\section{Discussions}\label{sec_discussions}
\subsection{Summary}
 
In this paper, we propose an extended compass model that hosts subsystem symmetries on each row and column. 
The single-ion anisotropy $D$ term offers extra tunability to the original compass model while preserving the subsystem symmetries, and leads to interesting physical consequences such as excitation immobility, critical Bose surface, and fracton-like excitations. 
Subsystem symmetries have been regarded indispensable to many interesting physical phenomena such as Bose metal and fracton topological order. 
We hope that our paper can shed light on experimental realization of subsystem symmetries in cold atom and condensed matter systems.

\subsection{Relevance to transition metal compounds}

Our extended compass model [Eq.~\eqref{Eq:ham}] has potential relevance to $3d$ transition metal compounds. 
Considering a layered perovskite structure where transition metals are arranged a layered square lattice, each transition metal ion is surrounded by a distorted octahedra of O$^{2-}$ (like La$_2$CuO$_4$).
We assume the $t_{2g}$ orbital is partially filled, \textit{e.g.}, with $t_{2g}^{1}$, $t_{2g}^{2}$, $t_{2g}^{4}$, $t_{2g}^{5}$ filling, so that the orbital angular momentum is active and can be described by an effective spin-1 operator.
The single-ion anisotropy $D$ term arises from the energy splitting between the $xy$ and $xz$/$yz$ orbitals. 
Without loss of generality we consider the $t_{2g}^{1}$ filling case, which is relevant to V$^{3+}$ or Ti$^{4+}$. The total orbital and spin angular momentum are defined as
\begin{equation}
\mathbf{L}=i\sum_{\sigma}\mathbf{d}_{\sigma}^{\dagger}\times\mathbf{d}_{\sigma},
\end{equation}
\begin{equation}
\mathbf{S}=\frac{1}{2}
\sum_{\alpha}d_{\alpha\sigma}^{\dagger}\boldsymbol{\sigma}_{\sigma\sigma'}d_{\alpha\sigma'},
\end{equation}
where we denote $d_{\sigma}=\left(d_{yz\sigma},d_{xz\sigma},d_{xy\sigma}\right)^{T}$.
The index $\alpha=xy,yz,xz$ sums over orbitals and $\sigma=\uparrow,\downarrow$ sums over spins. The effective orbital angular momentum $\boldsymbol{l}$ comes from projecting total orbital angular momentum $\mathbf{L}$ onto $t_{2g}$ manifold
\[
P_{t_{2g}}\mathbf{L}P_{t_{2g}}=-\boldsymbol{l}
\]
where $\boldsymbol l$ carries effective spin-1.

Then let us analyze the superexchange process between TM local moments. We first consider two TM local moments connected by a $x$~bond, with the O$^{2-}$ ligand located at the bond center. The hopping is dominated by 
\begin{equation}
H_{x,\text{hop}}=-t\sum_{i,j=i+\hat{e}_x,\sigma}\left(d_{i,xy\sigma}^{\dagger}d_{j,xy\sigma}+d_{i,xz\sigma}^{\dagger}d_{j,xz\sigma}+\text{H.c.}\right)
\end{equation}
We notice that only hopping between the same orbital species is allowed. The direct $\pi$-bond $d_{yz}\rightarrow d_{yz}$ hopping has negligible magnitude hence is not considered here. Similarly, the hopping on the $y$~bond reads
\begin{equation}
H_{y,\text{hop}}=-t\sum_{i,j=i+\hat{e}_y,\sigma}\left(d_{i,xy\sigma}^{\dagger}d_{j,xy\sigma}+d_{i,yz\sigma}^{\dagger}d_{j,yz\sigma}+\text{H.c.}\right)
\end{equation}
Hopping process beyond nearest neighbor is not considered due to the strong localization of $3d$ electrons. The spin-orbit coupling is also weak for $3d$ electrons hence is also ignored here.
The tetrahedral environment of the magnetic ion induce a small splitting $\Delta$ between $d_{xy}$ and $d_{xz}$/$d_{yz}$ orbitals,
\begin{equation}
H_{\Delta}=-\Delta\sum_{i,\sigma}d_{i,xy\sigma}^{\dagger}d_{i,xy\sigma}.
\end{equation}
The on-site interaction of $t_{2g}$ orbitals takes the complicated Kanamori form~\cite{kanamori}
\begin{align}
H_{\text{int}}=\sum_{i} & \big{[} \frac{U}{2}\sum_{\alpha}n_{i\alpha}^{2}+\sum_{\alpha>\beta}U'n_{i\alpha}n_{i\beta} \nonumber \\
&-2J_{H}(\mathbf{S}_{i\alpha}\cdot\mathbf{S}_{i\beta}+\frac{1}{4}n_{i\alpha}n_{i\beta}) \nonumber \\ 
& +J'(c_{i\beta\downarrow}^{\dagger}c_{i\beta\uparrow}^{\dagger}c_{i\alpha\uparrow}c_{i\alpha\downarrow}+\text{H.c.}) \big{]}
\end{align}
with $U=U'+J_{H}+J'$ and $J_H=J'$ for rotationally invariant orbitals.

The magnetic exchange of the full Hamiltonian 
$H_{\text{tot}}=H_{\text{int}}+H_\Delta+H_{x,\text{hop}}+H_{y,\text{hop}}$ 
can be obtained from second-order perturbation theory.
Following Kugel and Khomskii, the effective exchange interaction at the strong-coupling limit $U\gg t, J_H, |D|$ takes the form
\begin{align}
H_{\text{eff}} & =\sum_{i,j=i+\hat{e}_x}\frac{t^{2}}{4U}\left(1+4\mathbf{S}_{i}\cdot\mathbf{S}_{j}\right)\big{[}l_{i}^{x}l_{j}^{x}+\left(l_{i}^{x}\right)^{2}\left(l_{j}^{x}\right)^{2} \nonumber\\
&+q_{i}^{yz}q_{j}^{yz}+q_{i}^{y^{2}-z^{2}}q_{j}^{y^{2}-z^{2}}-\left(l_{i}^{x}\right)^{2}-\left(l_{j}^{x}\right)^{2}\big{]} \nonumber \\
 & +\sum_{i,j=i+\hat{e}_y}\frac{t^{2}}{4U}\left(1+4\mathbf{S}_{i}\cdot\mathbf{S}_{j}\right)\big{[}l_{i}^{y}l_{j}^{y}+\left(l_{i}^{y}\right)^{2}\left(l_{j}^{y}\right)^{2} \nonumber \\
&+q_{i}^{xz}q_{j}^{xz}+q_{i}^{x^{2}-z^{2}}q_{j}^{x^{2}-z^{2}}-\left(l_{i}^{y}\right)^{2}-\left(l_{j}^{y}\right)^{2}\big{]} \nonumber \\
 & -\Delta\sum_{i}\left(l_{i}^{z}\right)^{2}.
 \label{Eq:KK}
\end{align}
Here $q^{xz}=l^{x}l^{z}+l^{z}l^{x}$,$q^{yz}=l^{y}l^{z}+l^{z}l^{y}$, $q^{x^{2}-z^{2}}=\left(l^{x}\right)^{2}-\left(l^{z}\right)^{2}$, and $q^{y^{2}-z^{2}}=\left(l^{y}\right)^{2}-\left(l^{z}\right)^{2}$
are quadrupole operators of the orbital angular momentum. 
From the effective model we find that the system is invariant under the following subsystem symmetries, analogous to our extended compass model in the main text: for each row $r$ we define $\mathcal{P}_{r}$ as $\pi$ rotation about the $y$ axis acting on each orbital at the this row,
\begin{equation}
    \mathcal{P}_{r}=\prod_{i\in r}e^{i\pi l_{i}^{y}}.
    \label{Eq:P}
\end{equation}
Similarly, for each column $c$ we define $\mathcal{Q}_{c}$ as $\pi$ rotation
about the $x$ axis acting on each orbital at this column, 
\begin{equation}
    \mathcal{Q}_{c}=\prod_{i\in c}e^{i\pi l_{i}^{x}}.
    \label{Eq:Q}
\end{equation}

If we focus on the orbital degrees of freedom and perform a simple mean-field treatment on the spins, $\frac{t^{2}}{4U}\left(1+4\mathbf{S}_{i}\cdot\mathbf{S}_{j}\right)=J$,
we find that the orbital exchange part takes the following form:
\begin{align}
H_{\text{eff}}^{orb} & =\sum_{i,j=i+\hat{e}_x} \!J\! \left[l_{i}^{x}l_{j}^{x}
\!+\!\left(l_{i}^{x}\right)^{2}\left(l_{j}^{x}\right)^{2}
\!+\!q_{i}^{yz}q_{j}^{yz}
\!+\!q_{i}^{y^{2}\!-\!z^{2}}q_{j}^{y^{2}\!-\!z^{2}}\right]\nonumber \\
 & 
 +\sum_{i,j=i+\hat{e}_y}\!J\!\left[l_{i}^{y}l_{j}^{y}
\!+\!\left(l_{i}^{y}\right)^{2}\left(l_{j}^{y}\right)^{2}
\!+\!q_{i}^{xz}q_{j}^{xz}
\!+\!q_{i}^{x^{2}\!-\!z^{2}}q_{j}^{x^{2}\!-\!z^{2}}\right]\nonumber \\
 & -D\sum_{i}\left(l_{i}^{z}\right)^{2}
 \label{Eq:orb}
\end{align}
with $D=\Delta-2J$. 
This orbital model is very much like the extended compass model [Eq.~\eqref{Eq:ham}], with additional quadrupole-quadrupole interactions. 
However, such quadrupole-quadrupole interactions also preserve the subsystem symmetries [Eqs.~\eqref{Eq:P}~and~\eqref{Eq:Q}]. 
More properties associated with this model will be discussed in a follow-up paper~\cite{unpublished}.

\begin{acknowledgments}

We thank Rong Yu for valuable discussions. 
This work was supported by the National Key R\&D Program of China (Grant No. 2022YFA1403700), the National Natural Science Foundation of China (Grant No. 11925402), Guangdong Province (Grants No. 2020KCXTD001 and No. 2016ZT06D348), and the Science, Technology and Innovation Commission of Shenzhen Municipality (Grants No. ZDSYS20170303165926217, No. JAY20170412152620376, and No. KYTDPT20181011104202253). 
Z.L. gratefully acknowledge research support from the National Natural Science Foundation of China (Grants No. 12104313, No. 12034014), Shenzhen Natural Science Fund (the Stable Support Plan Program 20220810161616001) and Foundation from Department of Science and Technology of Guangdong Province (No. 2021QN02L820). 
C.-J.H. gratefully acknowledges research support from Gang Chen by the Ministry of Science and Technology of China with Grants No. 2021YFA1400300, the National Science Foundation of China with Grant No. 92065203, and the Research Grants Council of HongKong with Grant No. C7012-21GF. 
C.L. gratefully acknowledges research support by the National Natural Science Fund of China (Grant No. 12347101). 
The numerical simulations were supported by Center for Computational Science and Engineering of SUSTech and Tianhe-2.

\end{acknowledgments}

\appendix 
\counterwithout{equation}{section}               
\setcounter{equation}{0}                        
\renewcommand{\theequation}{A.\arabic{equation}} 

\section{\label{app:A}Effective Hamiltonian of the large-\texorpdfstring{$D$}{Lg} ferroquadrupole phase}
The effective Hamiltonian of the ferroquadrupole phase can be derived from the second-order perturbation theory in the large-$D$ limit. To perform the perturbation theory, we split the model [Eq.~\eqref{Eq:ham}] into two parts: $\mathcal H_0=-D\sum_\mathbf{r}(S_\mathbf{r}^z)^2$ as the the unperturbed part and $\mathcal{V}=\sum_{\mathbf{r}}J(S_{\mathbf{r}}^{x} S_{\mathbf{r}+\hat{e}_x}^{x} + S_{\mathbf{r}}^{y} S_{\mathbf{r}+\hat{e}_y}^{y})$ as the perturbation. The leading nontrivial contribution of $V$ appears at the second-order perturbation, which takes the form
\begin{align}
    H_{\text{eff}}^{(2)} = \mathscr{P}\mathcal{V}\mathscr{Q}\frac{1}{E_0-\mathscr{Q}\mathcal{H}_0\mathscr{Q}} \mathscr{Q}\mathcal{V}\mathscr{P},
\end{align}
where $\mathscr{P} = \otimes_\mathscr{r} \mathscr{P}_\mathbf{r}$ represents projection onto the ground-state manifold of $\mathcal{H}_0$, $\mathscr{Q}=1-\mathscr{P}$, and $E_0$ is the ground-state energy of $\mathcal{H}_0$. After straightforward calculations, the resulting effective theory turns out to be a ferromagnetic Ising model of the $\tau$ variables, as described in Eq.~\eqref{Eq:Heff2} in the main text.

It should be noted that the ferroquadrupolar ordering here arise purely from the perturbative effect, which involves mutual quantum entanglement between different sites. However, such mutual entanglement effect has been ignored in the semiclassical approximation as well as the linear flavor-wave theory, in which the quantum entanglement is restricted within each site. Therefore, the ferroquadrupolar order cannot be directly obtained in our semiclassical Monte Carlo simulations. To produce the correct results the second-order corrections, Eq.~\eqref{Eq:Heff2} must be explicitly added in the Hamiltonian in our calculations.

\section{\label{app:B}Sampled quantities in numerical simulations}

\begin{figure*}[!t]
    \includegraphics[width=\textwidth]{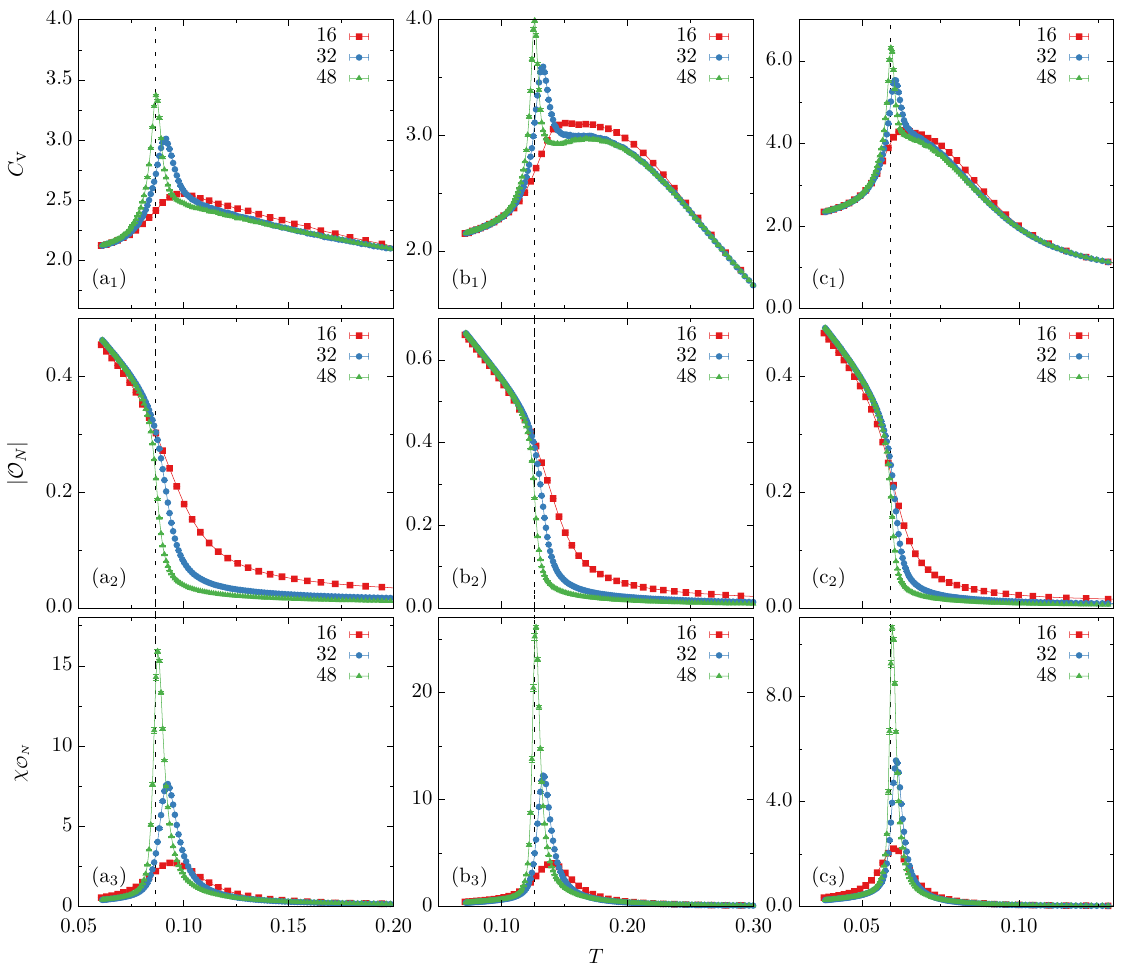}
    \caption{Temperature evolution of the specific head $\Cv$, the Ising nematic order $\mcO_N$ and its fluctuation $\chi_{\mcO_N}$ with different systems and $D$ values. The $D$ values of different columns are $-2.0$, $-0.5$, and $2.0$ sequentially. The dashed lines indicate the $\Cv$ peaks of the system size $L=48$, which are used as the phase boundary, the solid line, shown in Fig.~\ref{fig:2}.}
    \label{fig:transition}
\end{figure*}

\begin{figure*}[!t]
    \includegraphics[width=1.0\textwidth]{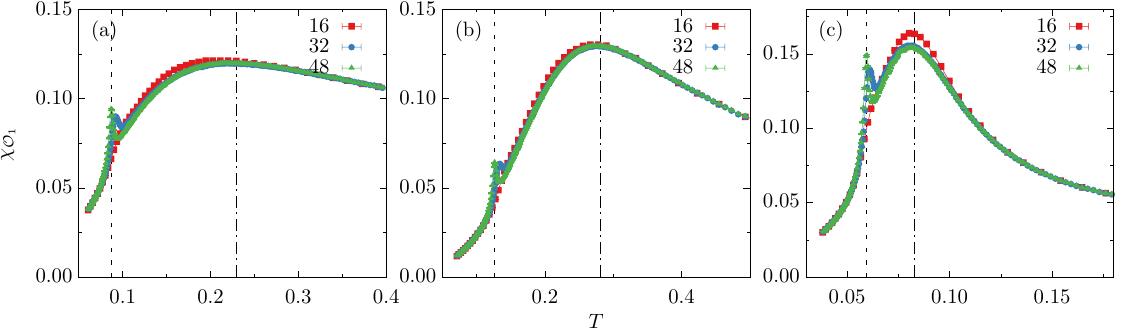}
    \caption{
    The fluctuation $\chi_{\mcO_1}$ of the 1D order over temperature. The $D$ values of (a), (b), and (c) are $-2.0$, $-0.5$, and $2.0$ respectively. 
    The dashed lines indicate the peak of the specific heat. 
    The dash-dotted lines indicate the maximum value of $\chi_{\mcO_1}$ of the system size $L=48$, which are used to determine the crossover, the blue-triangle points, shown in Fig.~\ref{fig:2}.}
    \label{fig:XO1}
\end{figure*}

\begin{figure*}[!t]
    \includegraphics[width=1.0\textwidth]{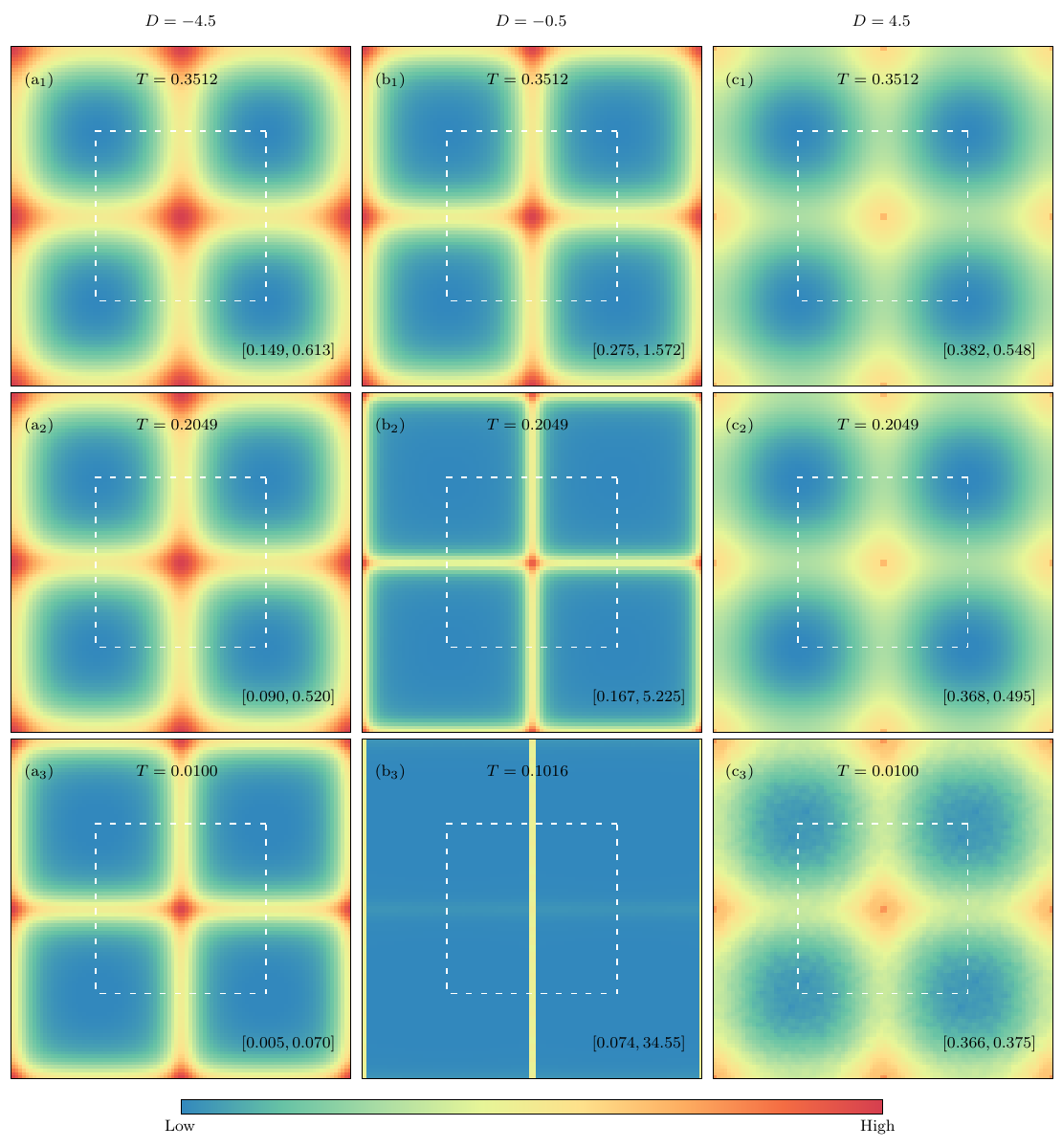}
    \caption{The temperature evolution of the SSF when $D=-4.5$, $-0.5$, $4.5$. The white-dashed line is the boundary of the first Brillouin zone.}
    \label{fig:ssf}
\end{figure*}

The symmetries of our model [Eq.~\eqref{Eq:ham}] have already been analyzed in the main text. However, for the sake of clarity and to provide a better understanding of why we are measuring certain observables, we will provide a brief analysis of these symmetries again.
As stated in the main text, there are Ising subsystem symmetries, which are described by operators $\mathcal{P}_j$ and $\mathcal{Q}_l$. According to Ref.~\cite{Ref_DimensionalReduction_00}, these one-dimensional subsystem symmetries will not be spontaneously broken at any finite temperatures. Besides, the system also host time-reversal symmetry $\Theta$ and a spin-orbit-coupled $C_4$ rotational symmetry. In the $D \rightarrow -\infty$ limit, the single-ion $D$ term dominates, and in the ground state, all spins are in a state of $S^z_{\bm{r}}=0$, which is called ``large-$D$ quantum paramagnetic'' phase. When $D \rightarrow +\infty$, a twofold ferroquadrupole order are predicted by degenerate perturbation theory. In the intermediate $D$ regime, in which region the sMCM simulation is mainly performed, the model enters an Ising nematic liquid which preserves time-reversal symmetry but breaks the $C_4$ symmetry down to the $C_2$ symmetry. 
With increasing temperature, the system first steps into an intermediate phase, called as a nodal-line spin liquid, where the $C_4$ symmetry is recovered. For detecting this transition, the Ising nematic order $\mcO_N$, the fluctuation of $\chi_{\mcO_N}$ and the specific heat $\Cv$ are measured in sMCM simulations. Analogous to the ``spiral spin liquid'', in the nodal-line spin liquid, the spin structural factors are peaked not in some momentum points but sharply peaked along the entire $k_x$ and $k_y$ axes. 
The difference is that the nodal-line spin liquid has no spiral textures and that the nodal-line degeneracy is protected by subsystem symmetries, while for spiral spin liquids the spiral degeneracy is accidental.
Further increasing the temperature, we find a big hump in the specific heat, which signals a crossover to the high-temperature paramagnetic phase~(see Fig.~\ref{fig:transition}).
From the spin structural factors (Fig.~\ref{fig:ssf}), we observe that the high-temperature paramagnetic phase shows quite diffusive feature, while the intermediate nodal-line spin liquid is sharply peaked along the $k_x$ and $k_y$ axes, indicating well-established 1D correlation.
As a result, the onset to the nodal-line spin liquid can be detected by measuring the fluctuations of the 1D correlation $\mcO_1$ along the $x$ and $y$ directions.

Here we give the definitions of observables mentioned above and other helpful quantities are given as well.
\begin{enumerate}
    \item Specific heat
    $$\Cv=\frac{1}{T^2}(\braket{\mathcal H^2}-\braket{\mathcal H}^2).$$

    \item The Ising nematic order parameter
        $$\mcO_N=\frac{1}{N}\sum_{\mathbf{r}} (S_{\mathbf{r}}^{x} S_{\mathbf{r}+\hat{e}_x}^{x}-S_{\mathbf{r}}^{y} S_{\mathbf{r}+\hat{e}_y}^{y}).$$

    \item Fluctuation of the Ising nematic order
    $$\chi_{\mcO_N}=\frac{1}{N}\left(\braket{\mcO_N^2}-\braket{\mcO_N}^2\right).$$

    \item The 1D spin correlation
    $$\mcO_{1}=\mcO_x+\mcO_y,\ \mcO_\alpha=\frac{1}{N}\sum_{\mathbf{r}} S_{\mathbf{r}}^{\alpha} S_{\mathbf{r}+\hat{e}_\alpha}^{\alpha},\ \alpha=x,y.$$
Here $\mcO_x$ and $\mcO_y$ measure the 1D spin correlations along the $x$ and $y$ directions, respectively.

    \item Fluctuation of the 1D spin correlation $$\chi_{\mathcal{O}_1}=\frac{1}{N}\left(\braket{\mcO_1^2}-\braket{\mcO_1}^2\right).$$

\end{enumerate}

\section{\label{app:C}More numerical results}

We simulate the model on square lattice with periodic boundary conditions and with the linear system sizes $L=16,32,48$. Except for the standard Metropolis update, the parallel tempering is employed to overcome the nonergodic problem. After thermalizing systems to equilibration, about $4\times 10^6$ independent samples are collected for each single-ion anisotropy $D$ and each temperature $T$.

\subsection{Thermal phase transition and crossover}

To detect the thermal phase transition of the Ising nematic liquid, the specific heat $\Cv$, the Ising nematic order $\mcO_N$ and the fluctuation $\chi_{\mcO_N}$ of the Ising nematic order $\mcO_N$ are measured, and results of several $D$ values are shown in Fig.~\ref{fig:transition}.

It is obvious that there is a divergent peak in the specific heat at low temperatures indicating a phase transition, and their locations are changed with varying $D$ values. Besides, the Ising nematic order $\mcO_N$ increases rapidly near the transition as temperature decreases. 
This suggests that the $C_4$ symmetry is broken to $C_2$ at low temperatures. 
What is more, the divergent peaks of $\chi_{\mcO_N}$ are consistent of the peaks of $\Cv$, which implies these are phase transition points further. The phase boundary of the Ising nematic spin liquid is identified as the locations of the peaks in the specific heat of the system with size $L=48$. These results are shown in Fig.~2 in the main text as indicated by a red-solid line.

Above the phase transition, the $C_4$ symmetry is recovered as the temperature increases and no more symmetry is expected to restore. 
As expected, in the specific heat data we observe no further phase transitions at higher temperatures.
Due to the presence of the 1D subsystem symmetries, the system tends to adopt a nodal-line spin liquid state with the spin structure factor sharply peaked along the entire $k_x$ and $k_y$ axes, where the sharp peaks indicate well-established 1D correlations. 
However, at sufficiently high temperatures, such 1D correlations must be melt by the strong thermal fluctuations. 
Therefore, we anticipate that there is a crossover between the nodal-line spin liquid and the high-temperature paramagnet. 
From the specific heat data Fig.~\ref{fig:transition}, we indeed find a big hump above the nematic transition that suggests such crossover.
To confirm that the crossover is associated with the lost of 1D correlations, we measure the fluctuation of the 1D spin correlation $\chi_{\mcO_1}$ as shown in Fig.~\ref{fig:XO1}.
We also find peak-hump structure, where the positions of the peak and hump basically coincides that in the specific heat $C_\mathrm{v}$ as indicated by dashed lines in Fig.~\ref{fig:XO1}.

\subsection{Spin structural factor}
To clarify the spin structure in different phases, we measure the spin structural factor~(SSF), which is defined by Eq.~(16). For the convenience of the reader, we repeat its definition here,
\begin{equation}
    \mathcal S(\mathbf{Q})=\frac {1}{N}\sum_{\mathbf{rr}'}\langle \mathbf{S}_\mathbf{r}\cdot \mathbf{S}_{\mathbf{r}'} \rangle e^{i\mathbf{Q}\cdot(\mathbf{r}-\mathbf{r}')}.
\end{equation}
In Fig.~\ref{fig:ssf}, it shows the evolution of SSF over temperature when $D=-4.5, -0.5, 4.5$. First we demonstrate that the 1D subsystem symmetry along the $x$ or $y$ axis always holds at all temperature. Secondly, as the temperature decreases, the width of SSF near the $k_x=0$ or $k_y=0$ axis becomes smaller, which means the 1D nematic order gradually becomes apparent. Noticeably, for $D=-0.5$ case, the $C_4$ symmetry breaks to the $C_2$ symmetry after the system enters the Ising nematic liquid at low temperature.

For the $D=-4.5,4.5$ cases, there is no apparent difference between the high-temperature and low-temperature SSFs. We also detect the specific heat and find no finite-temperature peaks indicating that there is no finite-temperature transition, which is consistent with our expectation.

\section{\label{app:D}Validity of excitation mobility with flavor-wave interactions}
In the main text, our central conclusion on the mobility of the flavor-wave excitations was derived with linear flavor-wave approximation, in which the effect of flavor-wave interaction is ignored.  
Here we show that our conclusion still hold in presence of  flavor-wave interactions.
We start with the complete flavor-wave Hamiltonian,
\begin{equation}
    \mathcal{H}_\textrm{Flavor wave}=\mathcal{H}_2+\mathcal{H}_{\textrm{int}}+\textrm{const.},
\end{equation}
where $\mathcal{H}_2$ is the two-body linear flavor-wave Hamiltonian taking the form of Eqs.~\eqref{Eq:Ham_QP} and \eqref{Eq:Ham_FQ} in the quantum paramagnet and the ferroquadrupole phases, respectively. $\mathcal{H}_{\textrm{int}}$ denotes the flavor-wave interaction that contains four-body and higher order terms. 
Note that both $\mathcal{H}_2$ and $\mathcal{H}_{\textrm{int}}$ respect Ising subsystem symmetries [Eqs.~\eqref{Eq:P_ham}~and~\eqref{Eq:Q_ham}], 
hence linear mixing between the (bare) single flavor-wave excitation and (bare) flavor-wave pair is not allowed.

We start with the quadratic limit $\mathcal{H}_{\textrm{int}}=0$, and then turn on $\mathcal{H}_{\textrm{int}}$ adiabatically. 
As the interactions are turned on, the wave functions of the single flavor-wave bosons will become dressed with interactions. However, the symmetry representations of the wave functions are not expected to have any sudden change during this adiabatic process. Therefore, the symmetry relation [Eqs.~\eqref{Eq:sym_QP}~and~\eqref{Eq:sym_FQ}] still holds, and that our conclusion about the mobility of flavor-wave excitations still remain valid.

\bibliography{paper.bib}


\end{document}